\def\NONANON{}
\documentclass[conference]{IEEEtran}

\usepackage{amsmath,amssymb,amsfonts,amsthm,booktabs,graphicx}
\usepackage{url}
\DeclareMathOperator*{\argmax}{arg\,max}

\newtheorem{theorem}{Theorem}
\newtheorem{proposition}[theorem]{Proposition}
\newtheorem{lemma}[theorem]{Lemma}
\newtheorem{corollary}[theorem]{Corollary}
\newtheorem{definition}[theorem]{Definition}
\newtheorem{example}[theorem]{Example}
\newtheorem{remark}[theorem]{Remark}
\newtheorem{lemmaA}{Lemma}

\newcommand{\Z}{\mathbb{Z}}
\newcommand{\E}{\mathbb{E}}
\newcommand{\Prob}{\mathbb{P}}
\newcommand{\ccrit}{c_{\mathrm{crit}}}

\newif\ifanon
\ifdefined\NONANON\anonfalse\else\anontrue\fi

\begin{document}

\title{Optimal Additive Integer Masks for One-Guess Vulnerability under an Absolute-Moment Budget}

\ifanon
\author{\IEEEauthorblockN{Anonymous submission}}
\else
\author{
\IEEEauthorblockN{Joonas P\"a\"akk\"onen}
\IEEEauthorblockA{Dalarna University, Sweden\\
Email: jpa@du.se}
}
\fi

\maketitle

\begin{abstract}
A defender releases an integer-valued secret \(X\) (a count, a discretised measurement, a clock value) as \(Y=X+N\), where \(N\) is an independent integer-valued mask whose expected absolute value, the expected distortion of the released value, is at most \(c\). An adversary who knows the prior of \(X\) and the law of \(N\) observes \(Y\) once and makes a single guess. We determine exactly how small the adversary's success probability, the posterior Bayes vulnerability, can be made, and which masks achieve it. For a secret supported on an interval of \(m\) integers we give the optimum and an explicit optimiser for the uniform prior; by the maximal-leakage formula, the same mask minimises the worst-case-prior one-guess leakage for every prior with that support. For a biased bit, the simplest prior a defender can exploit, we characterise the masks that leak nothing on average, give the critical budget at which this becomes possible, and determine the entire cost--vulnerability frontier. It is the piecewise-linear interpolation of an explicit sequence of points generated by cost-optimally shifted truncated geometric masks, with infinitely many affine pieces accumulating at the critical budget, where the frontier meets the prior vulnerability with zero slope. The proof rests on a layer-cake decomposition of an arbitrary mask into truncated geometrics, along which both vulnerability and cost are exactly additive. The results give exact, prior-aware design guidance for additive integer noise against one-guess adversaries.
\end{abstract}


\section{Introduction}
Adding independent noise to a value before it is released is one of the oldest and most widely deployed ways of limiting what an observer can infer. Integer-valued noise is the natural choice when the released value is itself an integer: counting queries in differential privacy are answered with two-sided geometric noise~\cite{GRS12} or discrete Gaussian noise~\cite{CKS20}, and timing channels have been mitigated by randomising the clocks available to processes~\cite{Hu91}. Such mechanisms are usually designed for a fixed privacy definition, most often differential privacy, and their noise law is chosen from a parametric family.

This paper asks a different, operational question. Suppose the adversary knows the distribution of the secret and the distribution of the noise, sees one noisy release, and wins if a single guess of the secret is correct. Among \emph{all} integer noise laws whose expected absolute value is at most a budget \(c\), which one minimises the adversary's success probability, and what is the minimum? The success probability of the best single guess is the posterior Bayes vulnerability of quantitative information flow (QIF)~\cite{Smith09,QIFbook}; equivalently, we minimise min-entropy leakage for a fixed prior. The cost \(\E|N|\) is the expected absolute distortion of the released value, the quantity a data user pays for the protection.

\subsection{Threat model}
\emph{System.} A secret \(X\) with finite support in \(\Z\) and known prior \(p\) is released once through the additive channel \(Y=X+N\), where \(N\) is drawn independently of \(X\) from a mask law \(q\) on \(\Z\) chosen by the defender.

\emph{Defender.} The defender chooses \(q\) before seeing \(X\), subject to the distortion budget \(\E_q|N|\le c\). The defender may know the prior \(p\) (prior-aware design) or only its support (worst-case-prior design).

\emph{Adversary.} The adversary knows \(p\) and \(q\) (no security through obscurity), observes the single release \(Y\), and outputs one guess \(\hat X\); it wins if \(\hat X=X\). Its optimal success probability is the posterior Bayes vulnerability \(V(X\mid Y)\) (Proposition~\ref{prop:map}). Without the release it would win with probability \(V(X)=\max_x p(x)\).

\emph{Out of scope.} Repeated releases of the same secret, adversaries with several guesses or other gain functions, and guarantees that hold for every pair of neighbouring inputs, such as differential privacy. Section~\ref{sec:design} explains why the optimal masks here are not differentially private.

\subsection{Contributions}
\begin{enumerate}
\item For a secret on an interval of \(m\) integers, the exact minimal vulnerability under the uniform prior and an explicit optimiser are determined (Theorem~\ref{thm:uniform}). Through the maximal-leakage formula~\cite{IWK20}, the same mask minimises the worst-case-prior (capacity) one-guess leakage for \emph{every} prior with that support (Section~\ref{sec:uniformsec}).
\item For a biased bit, all masks that produce no extra average guessing advantage are characterised, the exact critical budget and the strict gap below it are given, and the exact initial segment is determined (Theorems~\ref{thm:zero}, \ref{thm:ccrit} and~\ref{thm:initial}).
\item For every bias \(p\in(1/2,1)\) the complete cost--vulnerability frontier is determined (Theorem~\ref{thm:frontier}). It is the linear interpolation of explicit points \(Q_k\), \(k\ge1\), obtained from truncated geometric masks of length \(k\), and it has infinitely many affine pieces accumulating at the critical budget (Corollary~\ref{cor:pieces}). On every piece an optimiser is a mixture of two adjacent truncated geometrics (Corollary~\ref{cor:structure}). The first two pieces beyond the initial segment are also certified directly by dual multipliers (Corollary~\ref{cor:cells}).
\item The frontier shows that knowing the prior changes the optimal design qualitatively: a mask that is optimal against the worst-case prior can leave a biased bit strictly more exposed than the prior-aware optimum (Section~\ref{sec:design}).
\end{enumerate}

\subsection{Scope and positioning}
The model is deliberately minimal: one additive channel, one release, one guess, and one linear cost. We view this as the canonical noise-addition channel of QIF, and the point of the paper is that, for this channel, the optimal mechanism can be found \emph{exactly} over all noise laws, including infinite-support ones, rather than within a parametric family. The arguments are elementary (layer-cake decompositions, sumset bounds, convexity and linear duality), and we regard this as a feature: the optimisers are explicit and can be checked by hand. The results cover two priors in full, the uniform prior on an interval and the biased bit. The first also settles the worst-case prior for every prior on an interval. The second is the simplest prior that a defender can exploit, and it already exhibits an infinite piecewise-linear frontier. General priors on three or more points remain open (Section~\ref{sec:limits}).

\section{Related work}
\emph{Quantitative information flow.} Smith~\cite{Smith09} introduced Bayes vulnerability and min-entropy leakage as the measures appropriate to one-try attacks. Braun, Chatzikokolakis and Palamidessi~\cite{BCP09} compared additive and multiplicative one-try leakage. The \(g\)-leakage framework of Alvim et al.~\cite{ACPS12} generalises Bayes vulnerability through gain functions; Bayes vulnerability is the case of the identity gain function. Additive and multiplicative capacities were studied by Alvim et al.~\cite{ACMMPS14}, and the theory is presented in book form in~\cite{QIFbook}. Dodis, Ostrovsky, Reyzin and Smith~\cite{DORS08} defined average min-entropy, which equals \(-\log V(X\mid Y)\) for the posterior Bayes vulnerability used here.

\emph{Channel design.} Khouzani and Malacaria~\cite{KM16,KM17} study the design of channels that minimise leakage subject to hard feasibility constraints and soft expected-utility constraints. They identify settings in which a channel is optimal for every leakage measure simultaneously, show that such universal optimality fails in general, and prove that a broad class of leakage-minimisation problems is convex. Our problem is an instance of constrained channel design in which the channel is restricted to be additive and the constraint is an absolute-moment budget. The contribution here is the closed-form solution of this specific infinite-dimensional problem, not a general design method.

\emph{Maximal leakage and guessing.} Issa, Wagner and Kamath~\cite{IWK20} showed that maximal leakage for finite alphabets equals \(\log\sum_y\max_{x:p(x)>0}\Prob(Y=y\mid X=x)\), so maximal leakage is the logarithm of the one-guess advantage under the worst-case prior on the given support. Saeidian, Cervia, Oechtering and Skoglund~\cite{SCOS21} solved a privacy--utility problem under maximal leakage and expected Hamming distortion; there the mechanisms are not restricted to be additive and the cost is different. Asoodeh et al.~\cite{Asoodeh17} studied privacy-constrained estimation and guessing under a different formulation that does not specialise to additive integer masks with an absolute-moment budget.

\emph{Noise-adding mechanisms.} Optimal additive noise has been determined for differential privacy: Geng and Viswanath~\cite{GV16} for \(\varepsilon\)-differential privacy, and Geng, Ding, Guo and Kumar~\cite{GDGK19} for \((0,\delta)\)-differential privacy under \(\ell_1\) and \(\ell_2\) costs. Ghosh, Roughgarden and Sundararajan~\cite{GRS12} showed that the geometric mechanism is universally utility-maximising for counting queries under differential privacy, and Canonne, Kamath and Steinke~\cite{CKS20} analysed discrete Gaussian noise. These works constrain the mechanism by differential privacy, a worst-case guarantee over neighbouring inputs; we constrain the adversary's average one-guess success for a given prior. The relation between differential privacy and min-entropy leakage was studied by Alvim et al.~\cite{AACP11}.

\emph{Side-channel masking.} Masking in the side-channel sense, for example in the noisy-leakage model of Prouff and Rivain~\cite{PR13}, splits a secret into random shares, which is a different threat model. A targeted search of the QIF, maximal-leakage, differential-privacy-noise and guessing-with-noise literature did not locate prior solutions of the exact cost-constrained additive-integer problems studied here.

\section{Preliminaries}
Let \(X\) have finite support and law \(p\), and let \(N\) be independent with law \(q\) on \(\Z\). Set \(Y=X+N\).

\begin{proposition}[MAP rule]\label{prop:map}
The optimal one-guess success probability is
\[
V(X\mid Y)=\sum_{y\in\Z}\max_x p(x)q(y-x),
\]
attained by any \(g(y)\in\argmax_x p(x)q(y-x)\). In particular \(V(X\mid Y)\ge V(X)\).
\end{proposition}
\begin{proof}
A strategy \(g\) succeeds with probability \(\sum_y p(g(y))q(y-g(y))\); maximise each term separately. For any fixed \(x_0\), the sum is at least \(p(x_0)\).
\end{proof}

For a fixed secret distribution the optimisation is over all probability laws \(q\) on \(\Z\) with \(\E_q|N|\le c\). The optimal value is denoted \(V^*(c)\).

\section{Uniform interval secret}\label{sec:uniformsec}
Let \(X\) be uniform on \(I_m=\{0,\dots,m-1\}\). Then
\[
V(q)=\frac1m\sum_y\max_{x\in I_m}q(y-x).
\]

\begin{lemma}[Layer cake]\label{lem:layer}
Let \(h=\|q\|_\infty\) and \(B_t=\{y:q(y)>t\}\). Then \(mV(q)=\int_0^h|B_t+I_m|\,dt\).
\end{lemma}
\begin{proof}
For each \(s\), \(\max_{y\in s-I_m}q(y)=\int_0^h\mathbf 1[s\in B_t+I_m]\,dt\). Sum over \(s\) and apply Tonelli's theorem.
\end{proof}

\begin{lemma}[One-dimensional sumset]\label{lem:sumset}
For nonempty finite \(A\subset\Z\), \(|A+I_m|\ge|A|+m-1\). For \(m\ge2\), equality holds if and only if \(A\) is an interval of integers.
\end{lemma}
\begin{proof}
\(A+I_m\supseteq A\cup(\max A+I_m)\), and the two sets meet only in \(\max A\). If \(A\) is an interval the sumset is an interval of the stated cardinality. If consecutive elements of \(A\) differ by at least \(2\), a gap point appears in the sumset.
\end{proof}

\begin{lemma}[Cheapest mask of given peak]\label{lem:fill}
For \(h\in(0,1]\) let \(r\ge0\) satisfy \(1/(2r+3)<h\le1/(2r+1)\) and \(\rho=1-(2r+1)h\in[0,2h)\). Let \(q_h\) place mass \(h\) on each of \(\{-r,\dots,r\}\) and \(\rho/2\) on each of \(\pm(r+1)\). Then \(\E_q|N|\ge C(h):=\E_{q_h}|N|=(r+1)(1-(r+1)h)\) for every \(q\) with \(\|q\|_\infty\le h\).
\end{lemma}
\begin{proof}
Direct computation gives the cost formula. Let \(R=r+1\) and \(\Delta=q-q_h\). For \(|y|<R\) both factors \(|y|-R\le0\) and \(\Delta(y)\le0\), so their product is nonnegative. For \(|y|=R\) the term vanishes. For \(|y|>R\) both factors are nonnegative. Hence \(\sum(|y|-R)\Delta(y)\ge0\).
\end{proof}

\begin{theorem}[Uniform secret]\label{thm:uniform}
Let \(r\ge0\) satisfy \(r(r+1)/(2r+1)\le c\le(r+1)(r+2)/(2r+3)\). Set \(h^*(c)=(r+1-c)/(r+1)^2\). Then
\[
\min_{\E|N|\le c}V(q)=\frac{1+(m-1)h^*(c)}{m},
\]
attained by \(q_{h^*(c)}\).
\end{theorem}
\begin{proof}
Lemmas~\ref{lem:layer} and~\ref{lem:sumset} together with \(\int_0^h|B_t|\,dt=1\) give \(mV(q)\ge1+(m-1)\|q\|_\infty\). The cost constraint and Lemma~\ref{lem:fill} force \(\|q\|_\infty\ge h^*(c)\). The candidate has cost exactly \(c\) and interval superlevel sets, so equality holds throughout. The argument covers infinite-support competitors. For \(m=1\) or \(c=0\) one has \(V\equiv1\).
\end{proof}

The optimiser is not in general uniform on an interval. A continuous-noise analogue is treated elsewhere.

\paragraph{Security reading: the worst-case prior}
Theorem~\ref{thm:uniform} is not only a statement about one prior. For any prior \(\pi\) whose support is exactly \(I_m\), the maximal leakage of the channel \(x\mapsto x+N\) is \(\log\sum_y\max_{x\in I_m}q(y-x)\)~\cite{IWK20}, which depends on \(\pi\) only through its support and equals \(\log\bigl(m\,V(q)\bigr)\) with \(V(q)\) the uniform-prior vulnerability above. The same quantity is the multiplicative Bayes capacity of the channel, which upper-bounds the multiplicative Bayes leakage for every prior~\cite{ACMMPS14,QIFbook}. Hence the mask \(q_{h^*(c)}\) minimises the worst-case-prior one-guess leakage over all masks with \(\E|N|\le c\), and the minimum is \(\log\bigl(1+(m-1)h^*(c)\bigr)\). A defender who does not know the prior beyond its range can therefore use Theorem~\ref{thm:uniform} directly; a defender who knows the prior can do better, which is what the next section quantifies for the simplest non-uniform case.

\section{Biased bit}
Let \(\Prob(X=0)=p\in(1/2,1)\), \(\Prob(X=1)=1-p\), and \(a=(1-p)/p\in(0,1)\). Define
\[
V_p(q)=\sum_y\max\bigl\{p\,q(y),(1-p)q(y-1)\bigr\}
\]
and
\[
V_p^*(c)=\inf\{V_p(q):\E_q|N|\le c\}.
\]
The baseline is \(V(X)=p\). Equivalently,
\[
V_p(q)=p+\sum_y\max\bigl(0,(1-p)q(y-1)-p\,q(y)\bigr).
\]
For \(p=1/2\) the problem is the case \(m=2\) of Theorem~\ref{thm:uniform}.

\begin{theorem}[Zero extra average advantage]\label{thm:zero}
\(V_p(q)=p\) if and only if \(q(n)\ge a\,q(n-1)\) for every \(n\in\Z\). Such laws are precisely the mixtures \(q=\sum_L w_L G_L\), where \(G_L(n)=(1-a)a^{n-L}\mathbf 1_{n\ge L}\), \(w_L\ge0\) and \(\sum_L w_L=1\).
\end{theorem}
\begin{proof}
The first claim is immediate from the excess representation. Given the inequality, set \(d(n)=q(n)-a\,q(n-1)\ge0\). Iterating and using \(a^k q(n-k)\to0\) yields \(q(n)=\sum_{L\le n}d(L)a^{n-L}\). Summing over \(n\) gives \(\sum d(L)=1-a\), so the weights \(w_L=d(L)/(1-a)\) produce the claimed mixture. Conversely each \(G_L\) satisfies the inequality, which is preserved by mixtures.
\end{proof}

\begin{corollary}
No finite-support mask achieves \(V_p(q)=p\). If \(q(n_0)>0\) then the ratio inequality forces \(q(n_0+k)\ge a^k q(n_0)>0\) for all \(k\ge0\).
\end{corollary}

\begin{remark}
\(V_p(q)=p\) concerns the average only. For a pure geometric the posterior at the left endpoint equals \(1\), and \(Y\) is never independent of \(X\).
\end{remark}

\begin{theorem}[Critical cost]\label{thm:ccrit}
Let \(\ell^*=\#\{j\ge1:a^j>1/2\}=\lceil\log2/\log(1/a)\rceil-1\). Then
\[
c_{\mathrm{crit}}(p)=\ell^*+\frac{a(2a^{\ell^*}-1)}{1-a}=\min_{\ell\in\Z_{\ge0}}\Bigl(\ell+\frac{a(2a^\ell-1)}{1-a}\Bigr),
\]
the minimum being attained at \(\ell^*\), an integer median of the geometric (and also at \(\ell^*+1\) in the tie case \(a^{\ell^*+1}=1/2\)). Moreover \(V_p^*(c)=p\) for \(c\ge c_{\mathrm{crit}}(p)\) and \(V_p^*(c)>p\) for \(c<c_{\mathrm{crit}}(p)\).
\end{theorem}
\begin{proof}
By Theorem~\ref{thm:zero} and linearity the cost of a zero-advantage mask is at least the minimum over pure shifts. For positive shifts the cost is larger, so it suffices to minimise \(f(\ell)=\E_{G_{-\ell}}|N|\) over \(\ell\ge0\). The difference \(f(\ell+1)-f(\ell)=1-2a^{\ell+1}\) changes sign at the stated integer, with a tie exactly when equality holds.

Existence of minimisers: a minimising sequence with \(\E|N|\le c\) is tight by Markov's inequality. A diagonal argument yields a subsequence converging pointwise to some \(q\). Tightness implies \(\sum q=1\). Fatou's lemma gives \(\E_q|N|\le c\). Scheff\'e's lemma then upgrades the convergence to \(\ell^1\). The bound \(|V_p(q)-V_p(q')|\le\|q-q'\|_1\) gives continuity of the objective, so a minimiser exists. (The case \(c=0\) is covered by Theorem~\ref{thm:initial}.) If \(V_p^*(c)=p\) for some \(c<c_{\mathrm{crit}}(p)\), the minimiser would be a zero-advantage mask of cost strictly below \(c_{\mathrm{crit}}(p)\), a contradiction. For \(c\ge c_{\mathrm{crit}}(p)\) the geometric at the optimal shift is feasible.
\end{proof}

\begin{theorem}[Initial segment]\label{thm:initial}
For \(0\le c\le1-p\), \(V_p^*(c)=1-pc\), attained by \(q=(1-c)\delta_0+c\delta_1\). Moreover \(V_p(q)\ge1-p\,\E|N|\) for every \(q\).
\end{theorem}
\begin{proof}
Choosing the first term of the maximum for \(s\le0\) and the second for \(s\ge1\),
\begin{align*}
V_p(q)&\ge p\,\Prob(N\le0)+(1-p)\Prob(N\ge0)\\
&=1-p\,\Prob(N>0)-(1-p)\Prob(N<0)\\
&\ge1-p\,\Prob(N\ne0)\ge1-p\,\E|N|,
\end{align*}
using \(1-p<p\) and \(\mathbf 1[y\ne0]\le|y|\) on \(\Z\). The two-point mask has cost \(c\) and vulnerability \(1-pc\) on the stated interval.
\end{proof}

\subsection{Layer-cake decomposition into truncated geometrics}
The key technical tool is a layer-cake representation of an arbitrary mask \(q\) as a mixture of truncated geometrics \(T_{L,k}\) (Definition~\ref{def:Q} below; \(T_{L,\infty}=G_L\)) along which both vulnerability and cost are exactly additive. The decomposition slices the reweighted profile \(w(n)=q(n)a^{-n}\) horizontally. It is not a greedy peeling: \(V_p\) is only convex, and a greedy decomposition into truncated geometrics need not preserve it.

\begin{lemmaA}[Geometric layer cake]\label{lem:LC}
Let \(q\) be a probability law on \(\Z\) with \(\E_q|N|<\infty\). There is a probability measure \(\nu\) on \(\Z\times\{1,2,\dots,\infty\}\) such that
\begin{gather*}
q=\int T_{L,k}\,d\nu(L,k),\qquad V_p(q)=\int V_p(T_{L,k})\,d\nu,\\
\E_q|N|=\int \E_{T_{L,k}}|N|\,d\nu .
\end{gather*}
\end{lemmaA}
\begin{proof}
Put \(w(n)=q(n)a^{-n}\ge0\). Since \(q\le1\) and \(a^{-n}\to0\) as \(n\to-\infty\), we have \(w(n)\to0\) as \(n\to-\infty\). Hence for every \(t>0\) the level set \(\{w>t\}\) is bounded below, and it is a disjoint union of at most countably many maximal integer intervals (components) \(I=\{L,\dots,L+k-1\}\) with \(L\in\Z\) and \(k\in\{1,2,\dots,\infty\}\); components may be unbounded above, and the support of \(q\) may be unbounded in both directions. For a component \(I\) put \(g_I(n)=a^n\mathbf 1_I(n)\). Its mass is \(M_I=a^L(1-a^k)/(1-a)\) (\(=a^L/(1-a)\) if \(k=\infty\)), and \(g_I=M_I\,T_{L,k}\). By the layer-cake formula \(w(n)=\int_0^\infty\mathbf 1\{w(n)>t\}\,dt\),
\[
q(n)=a^n w(n)=\int_0^\infty\sum_{I}g_I(n)\,dt,
\]
where for each \(t\) the sum runs over the components of \(\{w>t\}\). Let \(\nu(\{(L,k)\})=\int_0^\infty M_{\{L,\dots,L+k-1\}}\,\mathbf 1\{\{L,\dots,L+k-1\}\text{ is a component of }\{w>t\}\}\,dt\) (for each \(I\), the set of levels \(t>0\) at which \(I\) is a component is an interval, hence measurable). Then the display reads \(q=\sum_{(L,k)}\nu(\{(L,k)\})\,T_{L,k}\), and summing over \(n\) (Tonelli; every term is nonnegative) gives \(\nu(\Z\times\{1,\dots,\infty\})=\sum_n q(n)=1\). Multiplying the display by \(|n|\) and summing over \(n\) (Tonelli again) gives the cost identity.

For the vulnerability, write \(V_p(r)=p\sum_n r(n)+p\sum_n\bigl(a\,r(n-1)-r(n)\bigr)_+\) for any nonnegative summable \(r\); this follows from \(\max\{p\,r(n),(1-p)r(n-1)\}=p\,r(n)+p\bigl(a\,r(n-1)-r(n)\bigr)_+\) and is positively homogeneous in \(r\). Now \(a\,q(n-1)-q(n)=a^n\bigl(w(n-1)-w(n)\bigr)\) and
\[
\bigl(w(n-1)-w(n)\bigr)_+=\int_0^\infty\mathbf 1\{w(n-1)>t\ge w(n)\}\,dt .
\]
The integrand equals \(1\) exactly when \(n-1\) is the right end of a finite component of \(\{w>t\}\). Hence
\[
\sum_n\bigl(a\,q(n-1)-q(n)\bigr)_+=\int_0^\infty\sum_{I\ \text{finite}}a^{L+k}\,dt .
\]
For a single component, \(\sum_n(a\,g_I(n-1)-g_I(n))_+\) equals \(a^{L+k}\) if \(I\) is finite (the only positive term is at \(n=L+k\)) and \(0\) if \(I\) is infinite. Combining this with the mass identity and positive homogeneity gives \(V_p(q)=\int_0^\infty\sum_I V_p(g_I)\,dt=\int_0^\infty\sum_I M_I V_p(T_{L,k})\,dt=\int V_p(T_{L,k})\,d\nu\).
\end{proof}

No additivity of \(V_p\) over arbitrary mixtures is used: by convexity, \(V_p\bigl(\sum_i\alpha_iT_i\bigr)\le\sum_i\alpha_iV_p(T_i)\) in general, often strictly. Equality holds for this particular decomposition because, for each level \(t\), distinct components of \(\{w>t\}\) are separated by at least one site, so their down-steps never cancel. Lemma~\ref{lem:LC} reduces the original infinite-dimensional optimisation to a convex combination of the elementary truncated-geometric problems solved next.

\subsection{The frontier}
\begin{proposition}[Monotonicity and convexity]\label{prop:convex}
The function \(V_p^*\) is nonincreasing and convex on \([0,\infty)\), and continuous on \([0,\infty)\).
\end{proposition}
\begin{proof}
If \(c\le c'\) the feasible set for \(c\) is contained in that for \(c'\), so \(V_p^*\) is nonincreasing. The function \(V_p\) is convex, since a maximum of linear functions is convex and sums of convex functions are convex. For convexity, take feasible masks \(q_0,q_1\) for budgets \(c_0,c_1\) and \(\theta\in[0,1]\). The mixture \(\theta q_0+(1-\theta)q_1\) is feasible for \(\theta c_0+(1-\theta)c_1\) and, by convexity of \(V_p\), has vulnerability at most the corresponding convex combination. Taking infima yields convexity of \(V_p^*\). A finite convex function on an open interval is continuous, so \(V_p^*\) is continuous on \((0,\infty)\). At \(0\), Theorem~\ref{thm:initial} gives \(V_p^*(c)=1-pc\) for \(0\le c\le1-p\), so \(V_p^*(c)\to1=V_p^*(0)\) as \(c\downarrow0\).
\end{proof}

\begin{definition}[Truncated geometrics and the points \(Q_k\)]\label{def:Q}
For \(L\in\Z\) and \(k\in\{1,2,\dots\}\) let
\[
T_{L,k}(s)=\frac{1-a}{1-a^k}\,a^{s-L}\,\mathbf 1\{L\le s\le L+k-1\},
\]
and put \(T_{L,\infty}=G_L\). For \(k\ge1\) and \(m\ge0\) write
\[
u_k=\frac{a^k}{1-a^k},\qquad A_m=\sum_{i=1}^m a^i=\frac{a(1-a^m)}{1-a},
\]
\[
m_k=\#\{1\le j\le k-1:\ 2a^j>1+a^k\},
\]
and define
\begin{align*}
V_k&=p+(2p-1)\,u_k,\\
C_k&=\frac{a}{1-a}+(m_k-2A_{m_k})+u_k\bigl(2(m_k-A_{m_k})-k\bigr),
\end{align*}
\(Q_k=(C_k,V_k)\), and \(Q_\infty=(\ccrit(p),p)\). An index \(k\) is a \emph{transition} if \(m_{k+1}=m_k+1\), and a \emph{tie} if \(2a^{m_k+1}=1+a^k\).
\end{definition}

For example \(Q_1=(0,1)\) is the point mass \(\delta_0\) and \(Q_2=(1-p,\,1-p+p^2)\).

\begin{theorem}[Biased-bit frontier]\label{thm:frontier}
Let \(p\in(1/2,1)\). Then \(0=C_1<C_2<C_3<\cdots\), \(C_k\to\ccrit(p)\), \(V_k\downarrow p\), and
\[
V_p^*(c)=\begin{cases}
V_k-\sigma_k\,(c-C_k), & C_k\le c\le C_{k+1},\ k\ge1,\\
p, & c\ge\ccrit(p),
\end{cases}
\]
with slopes
\begin{align*}
\sigma_k&=\frac{V_k-V_{k+1}}{C_{k+1}-C_k}=\frac{2p-1}{E_k},\\
E_k&=k-A_k-2(m_k-A_{m_k})-\theta_k(1-a^k),
\end{align*}
where
\[
\theta_k=\begin{cases}0,& m_{k+1}=m_k,\\[1mm] \dfrac{2a^{m_k+1}-1-a^{k+1}}{a^k(1-a)}\in(0,1],& m_{k+1}=m_k+1.\end{cases}
\]
The slopes satisfy \(\sigma_1\ge\sigma_2\ge\cdots>0\) and \(\sigma_k\to0\), and \(\sigma_{k-1}=\sigma_k\) holds if and only if \(k\) is a tie. For \(C_k\le c\le C_{k+1}\) an optimiser is the mixture
\[
q_c=\vartheta T_{-m_k,k}+(1-\vartheta)T_{-m_{k+1},k+1},\qquad
\vartheta=\frac{C_{k+1}-c}{C_{k+1}-C_k}.
\]
\end{theorem}

\begin{figure*}[t!]
\centering
\includegraphics[width=0.85\textwidth]{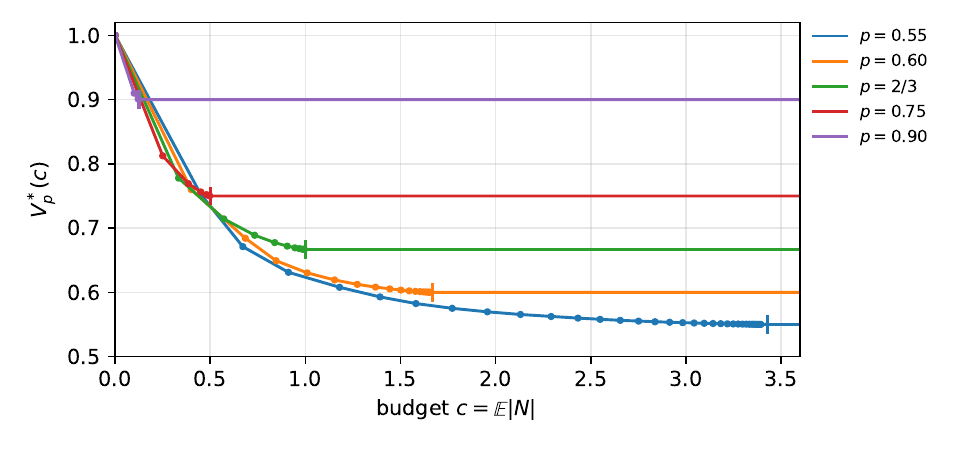}
\caption{Cost--vulnerability frontier for several biases. Dots mark the vertices \(Q_k\); vertical ticks mark the critical cost \(c_{\mathrm{crit}}(p)\), beyond which the frontier is constant at \(p\).}
\label{fig:frontier}
\end{figure*}

\begin{figure}[t]
\centering
\includegraphics[width=0.9\columnwidth]{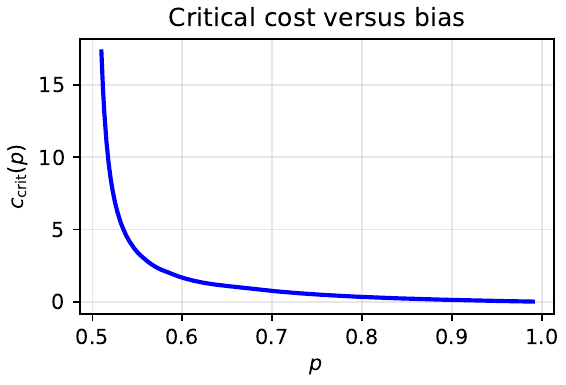}
\caption{Critical cost \(c_{\mathrm{crit}}(p)\) as a function of the bias \(p\in[0.51,0.99]\). The plot is restricted to \(p>1/2\) because \(c_{\mathrm{crit}}(p)\to\infty\) as \(p\downarrow1/2\).}
\label{fig:ccrit}
\end{figure}

\begin{figure}[t]
\centering
\includegraphics[width=0.9\columnwidth]{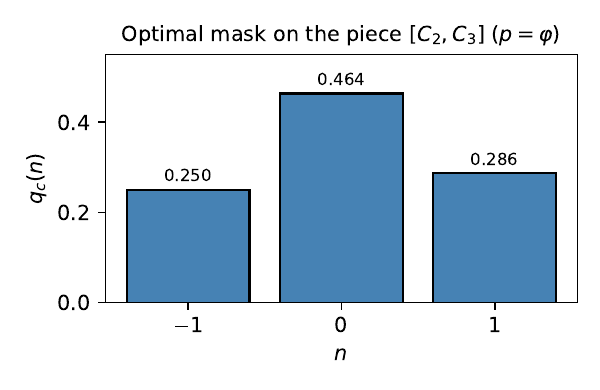}
\caption{Example optimal mask at the midpoint of the piece \([C_2,C_3]\) for \(p=\varphi=(\sqrt5-1)/2\approx0.618\) (so \(a=\varphi\)): the equal mixture of \(T_{0,2}\) and \(T_{-1,3}\). Here \(2a=1+a^3\), so \(k=3\) is a tie and the shift \(-1\) is as cheap as \(-m_3=0\) (Lemma~\ref{lem:T}(2)). Support is \(\{-1,0,1\}\) with masses approximately \(0.250\), \(0.464\), \(0.286\).}
\label{fig:optimiser}
\end{figure}

Figure~\ref{fig:frontier} shows the resulting frontier for several values of the bias. Figure~\ref{fig:ccrit} shows how the critical cost depends on the bias. Figure~\ref{fig:optimiser} shows a typical optimal mask on an intermediate piece.

\begin{example}[Numerical illustration]\label{ex:num}
For \(p=0.7\) one has \(a=3/7<1/2\), so \(\ell^*=0\) and \(c_{\mathrm{crit}}(0.7)=a/(1-a)=3/4\). The first points are \(Q_1=(0,1)\), \(Q_2=(3/10,79/100)\),
\begin{gather*}
Q_3=\bigl(\tfrac{39}{79},\tfrac{58}{79}\bigr)\approx(0.4937,0.7342),\\
Q_4=\bigl(\tfrac{177}{290},\tfrac{4141}{5800}\bigr)\approx(0.6103,0.7140).
\end{gather*}
On the piece \([C_2,C_3]\) the optimiser is the mixture of the optimally shifted length-2 and length-3 truncated geometrics with weights linear in \(c\).
\end{example}

\begin{lemma}[Properties of truncated geometrics]\label{lem:T}
(1) \(V_p(T_{L,k})=V_k\) for every \(L\) and every \(k\in\{1,2,\dots,\infty\}\).
(2) The minimal cost of a truncated geometric of length \(k\) is \(C_k\), attained by the shift \(L=-m_k\).
(3) \(m_k\) is nondecreasing in \(k\), increases by at most 1 at each step, and \(m_k\to\ell^*\).
See Appendix~\ref{app:T} for the proof.
\end{lemma}

\begin{lemma}[Convex position]\label{lem:convexpos}
Let \(E_k=(C_{k+1}-C_k)/(u_k-u_{k+1})\). Then \(E_k\) is given by the formula of Theorem~\ref{thm:frontier}, and \(E_1=1-a\). For \(k\ge2\) the increment \(E_k-E_{k-1}\) equals \(1-a^k\) if neither \(k-1\) nor \(k\) is a transition; \((1-\theta_k)(1-a^k)\) if \(k\) is a transition; and \(\bigl(2a^{m_k}-1-a^{k}\bigr)+\theta_{k-1}(1-a^{k-1})\) if \(k-1\) is a transition. In particular \(E_k\) is nondecreasing, \(E_k\ge1-a>0\), and \(E_k\to\infty\). Moreover \(E_k=E_{k-1}\) if and only if \(k\) is a tie.
See Appendix~\ref{app:convex} for the proof.
\end{lemma}

\begin{lemma}\label{lem:limit}
\(C_k\to\ccrit(p)\) and \(V_k\to p\).
\end{lemma}
\begin{proof}
\(V_k\to p\) because \(u_k\to0\). For every \(L\in\Z\) and \(k\ge1\), \(G_L=(1-a^k)T_{L,k}+a^kG_{L+k}\). With \(L=-\ell^*\) this gives \(C_k\le\E_{T_{-\ell^*,k}}|N|\le\ccrit(p)/(1-a^k)\). With \(L=-m_k\) and Theorem~\ref{thm:ccrit} it gives \(\ccrit(p)\le\E_{G_{-m_k}}|N|=(1-a^k)C_k+a^k\E_{G_{k-m_k}}|N|\le(1-a^k)C_k+a^k\bigl(k+a/(1-a)\bigr)\), since \(k-m_k\ge0\). Both bounds tend to \(\ccrit(p)\).
\end{proof}

\begin{proof}[Proof of Theorem~\ref{thm:frontier}]
Let \(g\) be the right-hand side of the theorem, with \(\sigma_k=(V_k-V_{k+1})/(C_{k+1}-C_k)\). Since \(V_k-V_{k+1}=(2p-1)(u_k-u_{k+1})\), we have \(\sigma_k=(2p-1)/E_k\). By Lemma~\ref{lem:convexpos}, \(C_k\) is strictly increasing, and \(\sigma_k\) is positive, nonincreasing and tends to \(0\). With Lemma~\ref{lem:limit}, \(g\) is well defined, continuous, nonincreasing and convex on \([0,\infty)\), since its slopes \(-\sigma_1\le-\sigma_2\le\dots\le0\) are nondecreasing. Also \(g(C_k)=V_k\) and \(g(\ccrit)=p\).

\emph{Lower bound.} Let \(\E_q|N|\le c\), and let \(\nu\) be the probability measure of Lemma~\ref{lem:LC}, so that \(V_p(q)=\int V_p(T_{L,k})\,d\nu\) and \(\E_q|N|=\int\E_{T_{L,k}}|N|\,d\nu\). Set \(C_\infty=\ccrit(p)\) and \(V_\infty=p\). By Lemma~\ref{lem:T} and Theorem~\ref{thm:ccrit}, \(V_p(T_{L,k})=V_k\) and \(\E_{T_{L,k}}|N|\ge C_k\) for all \(k\le\infty\). We use \(g(C_k)=V_k\), then Jensen's inequality for the convex function \(g\), then monotonicity of \(g\):
\begin{align*}
V_p(q)&=\int V_k\,d\nu=\int g(C_k)\,d\nu\\
&\ge g\Bigl(\int C_k\,d\nu\Bigr)\ \ge\ g\bigl(\E_q|N|\bigr)\ \ge\ g(c).
\end{align*}

\emph{Upper bound.} For \(C_k\le c\le C_{k+1}\), the mixture \(q_c\) of Lemma~\ref{lem:T}(2) has cost \(\vartheta C_k+(1-\vartheta)C_{k+1}=c\). By convexity of \(V_p\) and Lemma~\ref{lem:T}(1), \(V_p(q_c)\le\vartheta V_k+(1-\vartheta)V_{k+1}=g(c)\). For \(c\ge\ccrit(p)\), \(G_{-\ell^*}\) is feasible and has \(V_p=p\). Hence \(V_p^*=g\). The statements about the slopes are Lemma~\ref{lem:convexpos}.
\end{proof}

\begin{corollary}[Infinitely many pieces]\label{cor:pieces}
The pieces \([C_k,C_{k+1}]\) are infinitely many and accumulate only at \(\ccrit(p)\). The left derivative at \(\ccrit(p)\) is \(0\). The pieces \([C_{k-1},C_k]\) and \([C_k,C_{k+1}]\) have equal slopes if and only if \(k\) is a tie. There are at most \(\ell^*\) ties.
\end{corollary}
\begin{proof}
By Theorem~\ref{thm:frontier}, \(C_k\) increases strictly to \(\ccrit(p)\), which gives the first claim. The slope on \([C_k,C_{k+1}]\) is \(-\sigma_k\to0\), and \(V_p^*\) is convex, so its left derivative at \(\ccrit(p)\) is \(\lim_k(-\sigma_k)=0\). The third claim is Lemma~\ref{lem:convexpos}. Every tie is a transition (Appendix~\ref{app:convex}), \(m_1=0\), \(m_k\) increases by one at each transition, and \(m_k\le\ell^*\) for all \(k\) (Lemma~\ref{lem:T}(3)); hence there are at most \(\ell^*\) transitions and at most \(\ell^*\) ties.
\end{proof}

\begin{corollary}[Structure of optimisers]\label{cor:structure}
On each piece \([C_k,C_{k+1}]\) an optimiser is the mixture \(q_c\) of the two adjacent optimally shifted truncated geometrics \(T_{-m_k,k}\) and \(T_{-m_{k+1},k+1}\) of Theorem~\ref{thm:frontier}. If \(k\) (or \(k+1\)) is a tie, the shift \(-m_k-1\) (or \(-m_{k+1}-1\)) may be used instead.
\end{corollary}
\begin{proof}
The first claim is the upper-bound part of the proof of Theorem~\ref{thm:frontier}. At a tie the alternative shift has the same cost (Lemma~\ref{lem:T}(2)) and the same vulnerability (Lemma~\ref{lem:T}(1)), so the same argument applies.
\end{proof}

\begin{corollary}[Tail of the frontier]\label{cor:tail}
Let \(k_0=\min\{k\ge1:\ a^k<2a^{\ell^*}-1\}\). Then \(m_k=\ell^*\) for all \(k\ge k_0\), and for every \(k\ge k_0\)
\begin{align*}
C_k&=\frac{\ccrit(p)-a^k\bigl(k-\ell^*+a/(1-a)\bigr)}{1-a^k},\\
E_k&=k-\ccrit(p)-\ell^*+\frac{a^{k+1}}{1-a}.
\end{align*}
Consequently, for \(0<\varepsilon\le\sigma_{k_0}\), let \(k^*(\varepsilon)=\min\{k:\sigma_k<\varepsilon\}\) and \(x=(2p-1)/\varepsilon+\ccrit(p)+\ell^*\). Then \(x-a^{k^*+1}/(1-a)<k^*(\varepsilon)\le\lfloor x\rfloor+1\), with \(k^*(\varepsilon)=\lfloor x\rfloor+1\) whenever \(x-\lfloor x\rfloor\ge a^{\lfloor x\rfloor+1}/(1-a)\).
\end{corollary}
\begin{proof}
Since \(a^{\ell^*}>1/2\), we have \(2a^{\ell^*}-1\in(0,1]\), so \(k_0\) is finite. Recall \(m_k\le\ell^*\) and that the index set defining \(m_k\) is \(\{1,\dots,m_k\}\) (Lemma~\ref{lem:T}). For \(\ell^*\ge1\), \(m_k=\ell^*\) if and only if \(2a^{\ell^*}>1+a^k\) and \(\ell^*\le k-1\); the second condition follows from the first, because \(a^k<2a^{\ell^*}-1<a^{\ell^*}\). For \(\ell^*=0\), \(m_k=0\) and \(k_0=1\). In both cases \(m_k=\ell^*\) if and only if \(k\ge k_0\). Put \(\ell=\ell^*\) and \(b=a/(1-a)\). By Theorem~\ref{thm:ccrit}, \(\ccrit=\ell-b+2a^{\ell+1}/(1-a)\), and \(A_\ell=b-a^{\ell+1}/(1-a)\), so \(b+\ell-2A_\ell=\ccrit\) and \(2(\ell-A_\ell)=\ccrit+\ell-b\). Substituting these into Definition~\ref{def:Q} with \(m_k=\ell\) gives \((1-a^k)C_k=\ccrit-a^k(k-\ell+b)\). For \(k\ge k_0\) we also have \(m_{k+1}=\ell\), so \(\theta_k=0\). Theorem~\ref{thm:frontier} together with \(A_k=b-a^{k+1}/(1-a)\) then gives \(E_k=k-A_k-2(\ell-A_\ell)=k-\ccrit-\ell+a^{k+1}/(1-a)\).
For the second part, \(\sigma_k<\varepsilon\) if and only if \(E_k>(2p-1)/\varepsilon\). Since \(E_k\) is nondecreasing (Lemma~\ref{lem:convexpos}) and \(E_{k_0}\le(2p-1)/\varepsilon\), we get \(k^*(\varepsilon)=\min\{k\ge k_0:f(k)>x\}\), where \(f(k)=k+a^{k+1}/(1-a)\). The function \(f\) is strictly increasing, because \(f(k+1)-f(k)=1-a^{k+1}\). Moreover \(f(k)>k\), so \(k_0<x\), and \(f(\lfloor x\rfloor+1)>x\). This gives both bounds. If \(x-\lfloor x\rfloor\ge a^{\lfloor x\rfloor+1}/(1-a)\), then \(f(\lfloor x\rfloor)\le x\), and therefore \(k^*(\varepsilon)=\lfloor x\rfloor+1\).
\end{proof}
Equivalently, once \(m_k=\ell^*\), the optimal truncated geometric \(T_{-\ell^*,k}\) is the geometric \(G_{-\ell^*}\) of Theorem~\ref{thm:ccrit} conditioned on its first \(k\) atoms. The discarded tail has mass \(a^k\) and absolute moment \(a^k(k-\ell^*+a/(1-a))\). Since \(\sigma_k=(2p-1)/E_k\), slopes below \(\varepsilon\) are reached only after about \((2p-1)/\varepsilon+\ccrit(p)+\ell^*\) pieces. For \(\varepsilon=0.05\) one gets \(k^*=7,9,13\) for \(p=0.6,0.7,0.8\).

\subsection{The first pieces and explicit dual certificates}
\begin{corollary}[Certified cells]\label{cor:cells}
Let \(1/2<p\le\varphi=(\sqrt{5}-1)/2\), \(a=(1-p)/p\), and
\[
c_1=\frac{1+a^2}{1+a+a^2},\qquad c_2=\frac{1+a^2+2a^3}{1+a+a^2+a^3}.
\]
Then \(1-p<c_1<c_2\) and
\[
V_p^*(c)=\begin{cases}
p+\dfrac{(1-p)^2}{p}(1-c), & 1-p\le c\le c_1,\\[2mm]
1-p+p^3(2-c), & c_1\le c\le c_2.
\end{cases}
\]
On the first piece the optimiser is \(q(-1)=c-a(1-c)\), \(q(0)=1-c\), \(q(1)=a(1-c)\). On the second piece, with \(t=p^2(2-c)\), the optimiser is \(q(-1)=t\), \(q(0)=at\), \(q(1)=a^2t\), \(q(2)=1-t(1+a+a^2)\). Both affine functions are lower bounds on \(V_p^*(c)\) for every \(c\ge0\).
\end{corollary}
\begin{proof}
For any probability law \(q\) on \(\Z\) and any dual weights \(\alpha_y\in[0,1]\),
\begin{align*}
V_p(q)&\ge\sum_{y\in\Z}\bigl(\alpha_y p\,q(y)+(1-\alpha_y)(1-p)q(y-1)\bigr)\\
&=\sum_{n\in\Z}p\,\kappa(n)\,q(n),\qquad \kappa(n):=\alpha_n+a(1-\alpha_{n+1}).
\end{align*}
Since \(1/2<p\le\varphi\), we have \(a=(1-p)/p\in[\varphi,1)\), so \(a^2+a-1\ge0\), and direct algebra gives \(1-p<c_1<c_2\).

\emph{First piece.} Set \(\alpha_n=1\) for \(n\le0\), \(\alpha_1=1-a\), and \(\alpha_n=0\) for \(n\ge2\). Then \(\kappa(n)\) equals \(1\) for \(n\le-1\), \(1+a^2\) at \(n=0\), \(1\) at \(n=1\), and \(a\) for \(n\ge2\). Comparing with \(1+a^2(1-|n|)\) gives equality on \(\{-1,0,1\}\) and \(\ge\) everywhere else (at \(n\le-2\) trivially, since \(1>1-a^2\); at \(n\ge2\), \(a\ge 1-a^2\ge 1+a^2(1-n)\) since \(a^2+a-1\ge0\)). Hence \(V_p(q)\ge p+pa^2(1-\E_q|N|)=p+\frac{(1-p)^2}{p}(1-\E_q|N|)\) for every \(q\), proving the global lower bound. For \(1-p\le c\le c_1\), the stated three-point mask is nonnegative (since \(c\ge a/(1+a)=1-p\)), sums to \(1\), has cost \(c\), and satisfies \(a\,q(y-1)-q(y)\le0\) for all \(y\ne2\) (at \(y=0\) because \(c\le c_1\)), leaving only the excess term \((1-p)q(1)=\frac{(1-p)^2}{p}(1-c)\) at \(y=2\); thus it attains the bound.

\emph{Second piece.} Set \(\alpha_n=1\) for \(n\le-1\), \(\alpha_0=(2+a)p^2\), \(\alpha_1=p^2\), and \(\alpha_n=0\) for \(n\ge2\). Since \(p=1/(1+a)\) and \(a^2+a-1\ge0\), we have \(1-\alpha_0=(a^2+a-1)p^2\ge0\), so \(\alpha_n\in[0,1]\) for all \(n\). Direct evaluation gives \(\kappa(n)=a+p^2(2-|n|)\) on \(\{-1,0,1,2\}\) and \(\kappa(n)>a+p^2(2-|n|)\) for \(n\le-2\) (where \(\kappa(n)=1>a\)) and \(n\ge3\) (where \(\kappa(n)=a>a+p^2(2-n)\)). Multiplying by \(p\,q(n)\) and summing over \(n\) gives \(V_p(q)\ge 1-p+p^3(2-\E_q|N|)\) for every \(q\), proving the global lower bound. For \(c_1\le c\le c_2\), with \(t=p^2(2-c)\), the stated four-point mask has \(q(2)=1-t(1+a+a^2)\ge0\) (since \(c\ge c_1\)), sums to \(1\), has cost \(2-t(1+2a+a^2)=c\), and satisfies \(a\,q(1)-q(2)=t(1+a+a^2+a^3)-1\ge0\) (since \(c\le c_2\)). The excess terms at \(y=2\) and \(y=3\) sum to \(p(a^3t-q(2))+(1-p)q(2)=1-2p+pt\), yielding \(V_p(q)=1-p+p^3(2-c)\).
\end{proof}

\section{Design implications}\label{sec:design}
\emph{Using the frontier.} For a given budget \(c\) and bias \(p\), the defender identifies the piece \([C_k,C_{k+1}]\) containing \(c\) and deploys the corresponding mixture of two optimally shifted truncated geometrics (Theorem~\ref{thm:frontier}). The mixture weights are linear in \(c\). Returns diminish quickly: by Corollary~\ref{cor:tail}, the slope falls below \(\varepsilon\) only from about the piece \(k^*(\varepsilon)\approx(2p-1)/\varepsilon+\ccrit(p)+\ell^*\) onwards, where the optimisers have \(k^*+1\) support points. Near the critical budget the slope of the frontier vanishes: once \(c\) approaches \(\ccrit(p)\), further noise buys almost no reduction in the adversary's one-guess success.

\emph{Prior-aware versus worst-case design.} Let \(X\) be a biased bit and \(0<c\le 1-p\). The symmetric mask \(q_{h^*(c)}\) of Theorem~\ref{thm:uniform} with \(m=2\), which places \(1-c\) at \(0\) and \(c/2\) at \(\pm1\), minimises the maximal leakage \(\log\sum_y\max\{q(y),q(y-1)\}\) under the budget \(c\) (Section~\ref{sec:uniformsec}). A direct computation from the excess representation gives \(V_p(q_{h^*(c)})=1-c/2\), whereas the prior-aware optimum is \(V_p^*(c)=1-pc<1-c/2\) (Theorem~\ref{thm:initial}). The two-point optimiser \((1-c)\delta_0+c\delta_1\) of Theorem~\ref{thm:initial} also attains the minimal maximal leakage, because Theorem~\ref{thm:uniform} does not have a unique optimiser at this budget. Thus a mask chosen only to minimise worst-case-prior leakage can leave a known biased secret strictly more exposed than necessary, and the frontier of Theorem~\ref{thm:frontier} tells the defender which of the worst-case-optimal designs, if any, to prefer, and what to do beyond the initial segment.

\emph{Other leakage measures.} Mutual information penalises the whole posterior rather than the single best guess, and maximal leakage takes the worst case over priors. The exact frontier therefore supplies guidance that cannot be read off from either quantity.

\emph{No differential privacy.} The truncated geometric masks and the masks of Theorem~\ref{thm:uniform} have finite support, so an output near the edge of the support rules out some secret values. These masks give no \(\varepsilon\)-differential-privacy guarantee. This is consistent with the threat model: the guarantee here is about the average one-guess success of an adversary with a known prior, not about every pair of inputs.

\section{Limitations and future work}\label{sec:limits}
The threat model has one observation, one guess, a mask independent of the secret, and a pure absolute-moment cost. For the biased bit the frontier has been determined. Several questions remain open. First, uniqueness of the optimiser on intermediate pieces (apart from the degeneracy at ties). Second, prior-aware design for priors on three or more points beyond the uniform case. Third, other cost functions, other gain functions in the sense of \(g\)-leakage~\cite{ACPS12}, the modular analogue \(Y=X+N\bmod n\), multiple guesses, and repeated releases.

\section{Conclusion}
For the additive integer channel under an absolute-moment budget, the optimal masks against a one-guess adversary have been determined for a uniform interval secret, which also settles the worst-case prior on an interval, and, for a biased bit, along the entire cost--vulnerability frontier. The results supply exact, prior-aware design guidance for additive integer noise.

\ifanon
\section*{Acknowledgment}
AI tools (Grok, Grok Bot, and Gemini Neon) were used to draft and edit the text of all sections, to help develop and structure the proofs in Sections~III--V and the appendices, to search and check the literature cited in Section~II, and to write the scripts behind Figs.~1--3 and the exact-arithmetic checks. This includes the layer-cake proof of Lemma~\ref{lem:LC}, the proofs of Lemma~\ref{lem:T}(3), Corollaries~\ref{cor:pieces}--\ref{cor:cells}, the case analysis in Appendix~\ref{app:convex}, and a script (included in the supplementary material) that recomputes every numerical value quoted in the paper (Example~\ref{ex:num}, Fig.~\ref{fig:optimiser}, Corollaries~\ref{cor:tail}--\ref{cor:cells} and the table in Appendix~\ref{app:convex}) and checks the dual certificates of Corollary~\ref{cor:cells}. The author remains fully responsible for the accuracy of all statements, proofs, and numerical results.
\else
\section*{Acknowledgment}
AI tools (Grok, Grok Bot, and Gemini Neon) were used to draft and edit the text of all sections, to help develop and structure the proofs in Sections~III--V and the appendices, to search and check the literature cited in Section~II, and to write the scripts behind Figs.~1--3 and the exact-arithmetic checks. The same tools were used to prepare this version, including the layer-cake proof of Lemma~\ref{lem:LC}, which replaces the greedy peeling argument of earlier versions, the proofs of Lemma~\ref{lem:T}(3), Corollaries~\ref{cor:pieces}--\ref{cor:cells}, the case analysis in Appendix~\ref{app:convex}, and a script that recomputes every numerical value quoted in the paper and checks the dual certificates of Corollary~\ref{cor:cells}. Verification scripts are available at \url{https://github.com/fv2fm2z7rk-droid/additive-integer-masks-qif}. The author remains fully responsible for the accuracy of all statements, proofs, and numerical results.
\fi

\bibliographystyle{IEEEtran}
\bibliography{references}

\appendices
\section{Proof of the truncated-geometric properties}
\label{app:T}

\begin{proof}[Proof of Lemma~\ref{lem:T}]
(1) Write \(T=T_{L,k}\). By the excess representation,
\[
V_p(T)=p+\sum_y\max\bigl(0,(1-p)T(y-1)-p\,T(y)\bigr).
\]
On the support \(\{L,\dots,L+k-1\}\) one has \(T(y)=\gamma_k\,a^{y-L}\) with normalising constant \(\gamma_k=(1-a)/(1-a^k)\). At \(y=L\) the bracket is \((1-p)\cdot0-p\gamma_k<0\), so there is no excess. For \(L<y\le L+k-1\) it equals \(\gamma_k a^{y-1-L}\bigl((1-p)-pa\bigr)=0\), because \(pa=1-p\). Outside \(\{L,\dots,L+k\}\) both terms vanish. The only positive excess therefore occurs just beyond the right endpoint, at \(y=L+k\), where it equals \((1-p)\gamma_k a^{k-1}-p\cdot0=(1-p)\gamma_k a^{k-1}\). Since \((1-p)(1-a)a^{k-1}=(2p-1)a^k\) (use \(1-a=(2p-1)/p\) and \((1-p)/p=a\)), summing the single excess term yields
\[
V_p(T)=p+(1-p)\gamma_k a^{k-1}=p+(2p-1)u_k=V_k,
\]
independent of \(L\). For \(k=\infty\) the same computation applies with \(\gamma_\infty=1-a\); there is no right endpoint, so the excess vanishes and \(V_p(G_L)=p=V_\infty\) for every \(L\).

(2) The absolute-moment cost of a shift-\(L\) truncated geometric is
\[
\E|T_{L,k}|=\sum_{j=0}^{k-1}|L+j|\,\gamma_k\,a^j,\qquad \gamma_k=\frac{1-a}{1-a^k}.
\]
The first difference \(\Delta(L)=\E|T_{L+1,k}|-\E|T_{L,k}|\) equals
\[
\Delta(L)=\gamma_k\sum_{j=0}^{k-1}\operatorname{sign}(L+j+1/2)\,a^j.
\]
For \(L=-m\) with \(0\le m\le k\), exactly the indices \(j<m\) sit at negative positions, so
\[
\Delta(-m)=\gamma_k\Bigl(\sum_{j=m}^{k-1}a^j-\sum_{j=0}^{m-1}a^j\Bigr)=\frac{\gamma_k}{1-a}\bigl(2a^m-1-a^k\bigr).
\]
Thus \(\Delta(-m)>0\) if and only if \(2a^m>1+a^k\), i.e., if and only if the mass of \(T_{-m,k}\) at negative positions is \emph{strictly} less than half. Since \(2a^j\) is decreasing in \(j\), the set in the definition of \(m_k\) is \(\{1,\dots,m_k\}\); together with \(2a^0=2>1+a^k\) this gives \(2a^{m_k}>1+a^k\) and \(2a^{m_k+1}\le1+a^k\) (for \(m_k=k-1\) the latter reads \(2a^k\le1+a^k\), which always holds). Hence \(\Delta(L)>0\) for \(L\ge-m_k\) (for \(L\ge0\) all mass is at nonnegative positions), \(\Delta(-m_k-1)\le0\), and \(\Delta(L)<0\) for \(L\le-m_k-2\) (for \(L\le-k\) all mass is at negative positions). The cost is therefore minimised at \(L=-m_k\). The minimiser is unique unless \(k\) is a tie, \(2a^{m_k+1}=1+a^k\); then \(\Delta(-m_k-1)=0\) and \(L=-m_k-1\) attains the same cost \(C_k\). Substituting \(L=-m_k\) and summing the two finite geometric series (positive and negative parts) produces the closed form
\[
C_k=\frac{a}{1-a}+(m_k-2A_{m_k})+u_k\bigl(2(m_k-A_{m_k})-k\bigr).
\]

(3) By the proof of (2), the index set defining \(m_k\) is \(\{1,\dots,m_k\}\) and \(2a^{m_k+1}\le1+a^k\).
\emph{Monotonicity.} If \(1\le j\le k-1\) and \(2a^j>1+a^k\), then \(2a^j>1+a^{k+1}\) and \(j\le k\); hence \(m_{k+1}\ge m_k\).
\emph{Steps of at most one.} \(2a^{m_k+2}=a\cdot2a^{m_k+1}\le a+a^{k+1}<1+a^{k+1}\), so \(m_k+2\) does not belong to the index set for \(k+1\), and \(m_{k+1}\le m_k+1\).
\emph{Limit.} If \(2a^j>1+a^k\) then \(a^j>1/2\), so \(m_k\le\#\{j\ge1:a^j>1/2\}=\ell^*\) for every \(k\). Conversely, each \(j\le\ell^*\) has \(2a^j>1\), hence \(2a^j>1+a^k\) and \(j\le k-1\) for all sufficiently large \(k\). Thus \(m_k=\ell^*\) for all large \(k\).
\end{proof}

\section{Proof of the convex-position lemma}
\label{app:convex}

\begin{proof}[Proof of Lemma~\ref{lem:convexpos}]
Write \(C(k,m)=\E|T_{-m,k}|=\frac{a}{1-a}+(m-2A_m)+u_k(2(m-A_m)-k)\). Using \(u_k-u_{k+1}=a^k(1-a)/((1-a^k)(1-a^{k+1}))\) and \(u_{k+1}/(u_k-u_{k+1})=A_k\), subtracting at fixed shift \(m=m_k\) gives
\begin{align*}
&C(k+1,m_k)-C(k,m_k)\\
&\quad=(u_{k+1}-u_k)\bigl(2(m_k-A_{m_k})-k\bigr)-u_{k+1}\\
&\quad=(u_k-u_{k+1})\bigl(k-A_k-2(m_k-A_{m_k})\bigr).
\end{align*}
If \(m_{k+1}=m_k\), this equals \(C_{k+1}-C_k\). If \(m_{k+1}=m_k+1\), we add the shift step from the proof of Lemma~\ref{lem:T}(2):
\begin{align*}
&C(k+1,m_k+1)-C(k+1,m_k)\\
&\quad=-\Delta_{k+1}(-m_k-1)=-\frac{2a^{m_k+1}-1-a^{k+1}}{1-a^{k+1}}\\
&\quad=-(u_k-u_{k+1})\theta_k(1-a^k),
\end{align*}
where \(\Delta_{k+1}\) is the first difference \(\Delta\) of Appendix~\ref{app:T} for length \(k+1\). In both cases
\[
C_{k+1}-C_k=(u_k-u_{k+1})E_k
\]
with the expression for \(E_k\) given in Theorem~\ref{thm:frontier} (and at a transition, \(1+a^{k+1}<2a^{m_k+1}\le1+a^k\) gives \(\theta_k\in(0,1]\)). At \(k=1\): \(m_1=m_2=0\) (because \(2a<1+a^2\)), so \(\theta_1=0\) and \(E_1=1-a\).

\emph{Two facts about transitions.} (i) \emph{Indices \(k-1\) and \(k\) are never both transitions} (\(k\ge2\)). If \(k-1\) is a transition, then \(m_k=m_{k-1}+1\), and \(2a^{m_{k-1}+1}\le1+a^{k-1}\) (proof of Lemma~\ref{lem:T}(2)) gives
\[
2a^{m_k+1}=a\cdot2a^{m_k}\le a+a^k<1+a^{k+1},
\]
because \((1+a^{k+1})-(a+a^k)=(1-a)(1-a^k)>0\). Hence \(m_{k+1}=m_k\). (ii) \emph{Every tie is a transition.} If \(2a^{m_k+1}=1+a^k\), then \(m_k+1\le k\) and \(2a^{m_k+1}>1+a^{k+1}\), so \(m_{k+1}=m_k+1\).

By Lemma~\ref{lem:T}(3) and (i), for \(k\ge2\) exactly one of the following three cases occurs.
\begin{itemize}
\item \emph{Neither \(k-1\) nor \(k\) is a transition}: \(m_{k+1}=m_k=m_{k-1}\). The difference of the closed forms simplifies to
  \[
  E_k-E_{k-1}=1-a^k>0.
  \]
\item \emph{\(k\) is a transition}: \(m_{k+1}=m_k+1\) and \(m_k=m_{k-1}\). The extra median term contributes a factor \(\theta_k\), and
  \[
  E_k-E_{k-1}=(1-\theta_k)(1-a^k)\ge0,
  \]
  with equality if and only if \(\theta_k=1\), i.e., \(2a^{m_k+1}=1+a^k\): \(k\) is a tie.
\item \emph{\(k-1\) is a transition}: \(m_k=m_{k-1}+1\ge1\) and \(m_{k+1}=m_k\), and
  \[
  E_k-E_{k-1}=\bigl(2a^{m_k}-1-a^k\bigr)+\theta_{k-1}(1-a^{k-1})>0,
  \]
  since \(m_k\) belongs to the index set defining \(m_k\), i.e., \(2a^{m_k}>1+a^k\), and \(\theta_{k-1}>0\).
\end{itemize}
Hence \(E_k\ge E_{k-1}\), with equality if and only if \(k\) is a tie (by (ii), a tie at \(k\) falls in the second case). Thus \(E_k\) is nondecreasing and \(E_k\ge E_1=1-a>0\). By (i) and Lemma~\ref{lem:T}(3) there are at most \(\ell^*\) transitions, so for all large \(k\) the first case applies and \(E_k-E_{k-1}=1-a^k\ge1-a\); hence \(E_k\to\infty\). The slope claims follow at once from \(\sigma_k=(2p-1)/E_k\).
\end{proof}

\medskip\noindent\emph{Numerical values of the first points.} \(C_k\) and \(V_k\) for the cost-optimally shifted truncated geometric of length \(k\).
\begin{center}
\begin{tabular}{c|cc|cc}
\(k\) & \(C_k\) (\(p=0.6\)) & \(V_k\) & \(C_k\) (\(p=0.8\)) & \(V_k\) \\
\hline
1 & 0 & 1 & 0 & 1 \\
2 & 0.4000 & 0.7600 & 0.2000 & 0.8400 \\
3 & 0.6842 & 0.6842 & 0.2857 & 0.8095 \\
4 & 0.8462 & 0.6492 & 0.3176 & 0.8024 \\
5 & 1.0095 & 0.6303 & 0.3284 & 0.8006 \\
\end{tabular}
\end{center}
The exact values are the following rationals (\(a=2/3\) and \(a=1/4\)); the table above shows them rounded to four decimals, and each agrees with a brute-force minimisation of the cost over all shifts.
\begin{center}
\begin{tabular}{c|cc|cc}
\(k\) & \(C_k\) (\(p=0.6\)) & \(V_k\) & \(C_k\) (\(p=0.8\)) & \(V_k\) \\
\hline
2 & \(2/5\) & \(19/25\) & \(1/5\) & \(21/25\) \\
3 & \(13/19\) & \(13/19\) & \(2/7\) & \(17/21\) \\
4 & \(11/13\) & \(211/325\) & \(27/85\) & \(341/425\) \\
5 & \(213/211\) & \(133/211\) & \(112/341\) & \(273/341\) \\
\end{tabular}
\end{center}

\end{document}